\documentclass[twocolumn]{aastex631}

\usepackage{amsmath}
\usepackage{comment}

\newcommand{\arl}{\alpha_{\rm RL}}

\newcommand{\amt}{\alpha_{\rm MT}}
\newcommand{\am}{\alpha_{\rm M}}

\newcommand{\MBH}{M_{\bullet}}
\newcommand{\rtidal}{r_{\rm tidal}}
\newcommand{\rh}{r_{h}}
\newcommand{\rtidalb}{r_{\rm tidal,b}}
\newcommand{\tgw}{\tau_{\rm GW}}
\newcommand{\tkh}{\tau_{\rm KH}}
\newcommand{\tapp}{\tau_{\rm app}}
\newcommand{\ttb}{\tau_{\rm 2B}}
\newcommand{\tdyn}{\tau_{\rm dyn}}

\newcommand{\pfrac}[2]{\left( \frac{#1}{#2} \right)}

\begin{document}

\title{Unstable Mass Transfer from a Main-Sequence Star to a Supermassive Black Hole \\ and \\ Quasi-Periodic Eruptions}

\author[0000-0002-8304-1988]{Itai Linial}
\affiliation{Racah Institute, Hebrew University of Jerusalem, Jerusalem, 91904, Israel}
\affiliation{Institute for Advanced Study, 1 Einstein Drive, Princeton, New Jersey, 08540, USA}

\author[0000-0002-1084-3656]{Re'em Sari}
\affiliation{Racah Institute, Hebrew University of Jerusalem, Jerusalem, 91904, Israel}

\correspondingauthor{Itai Linial}
\email{itai.linial@mail.huji.ac.il,itailin@ias.edu}

\begin{abstract}

\textcolor{black}{We discuss the formation and evolution of systems comprised of a low-mass ($M_\star \lesssim 4 \, \rm M_\odot$) main sequence star, orbiting a $10^5-10^7 \, \rm M_\odot$ supermassive black hole with an orbital period of order $\sim$hours, and a mild eccentricity ($e\approx0.1-0.2$), episodically shedding mass at each pericenter passage. We argue that the resulting mass transfer is likely unstable, with Roche lobe overflow initially driven by gravitational wave emission, but then being accelerated by the star's expansion in response to its mass loss, undergoing a runaway process. We show that such systems are naturally produced by two-body gravitational encounters within the inner parsec of a galaxy, followed by gravitational wave circularization and inspiral from initially highly eccentric orbits. We argue that such systems can produce recurring flares similar to the recently identified class of X-ray transients known as Quasi-Periodic Eruptions, observed at the centers of a few distant galaxies.}

\end{abstract}


\section{Introduction}
Almost all galaxies harbor a supermassive black hole (SMBH) located at their center, surrounded by a dense nuclear star cluster, containing millions to billions of stars packed into a volume of a few cubic parsec \citep[e.g.,][]{Tremaine_2002,Volonteri_2021}.

These dense stellar environments are fertile grounds for a variety of dynamical processes, including the tidal disruption of stars as they pass nearby the central black hole \citep{Rees_1988}, the disintegration of stellar binaries and the subsequent expulsion of hypervelocity stars \citep{Hills_1988}, as well the formation of extreme-mass ratio inspirals (EMRIs) of compact objects towards the SMBH through \textcolor{black}{the emission of gravitational waves (GWs)} \citep{Amaro_Seoane_2018}.

Over the past decades, observations have revealed a plethora of transient phenomena associated with otherwise dormant galactic nuclei across the entire electromagnetic spectrum, and more recently even in neutrino. An important class of nuclear transients are tidal disrutption events (TDEs), which are now being routinely observed thanks to wide-field transient surveys such as ZTF, ASASSN, PAN-STARRS and in the near future, the Vera Rubin Observatory \citep[for a recent review, see][]{Gezari_2021}. These events occur when a star passes close to the SMBH and is torn apart by tidal forces, leading to high rates of infalling material, generating bright panchromatic flares. 

A related, but not as well-studied phenomena involves the inspiral of a star towards the SMBH through GW emission, until mass is gradually stripped from the star towards the black hole \citep{Dai_Blandford_2013,Linial_Sari_2017}. We refer to these systems as ``stellar-inspirals" or ``stellar-EMRIs" (to distinguish them from a compact object EMRI often discussed in the literature, e.g., \citealt{Amaro_Seoane_2018}). In this paper, we consider dynamical channels that bring stars to tight orbits around the SMBH, with hour-day periods. On these tight orbits, tides overcome the star's self gravity around pericenter, stripping a small fraction of its mass once an orbit, powering a bright, highly variable, electromagnetic emission. We suggest that these flares may have already been observed as Quasi-Periodic Eruptions (QPEs) - a recently identified class of bright soft X-ray flares associated with galactic nuclei \citep{Miniutti_2019,Giustini_2019,Arcodia_2021,Chakraborty_2021}.

\textcolor{black}{The structure of the paper is as follows. We introduce the timescales governing the relevant dynamical processes within the nucleus of a galaxy in section \ref{sec:Timescales}. The formation channels that place stars on orbits that tidally interact with the SMBH are then described in section \ref{sec:StellarInspiralProduction}. In section \ref{sec:InspiralFate}, the fate of stellar inspirals and the corresponding mass transfer phase is discussed. Section \ref{sec:QPEs_from_unstable_MT} is dedicated to the connection between QPEs and stellar inspirals, tying together the observables with our model. Our summary, conclusions and further discussion appear in section \ref{sec:Summary}.}

Shortly before the completion of this paper, we became aware of a recent work by \cite{Lu_Quataert_2022}, who have similarly considered a main-sequeunce star around an SMBH, undergoing unstable mass transfer as the origin of QPEs.

\section{Governing timescales} \label{sec:Timescales}

Consider a cluster of stars engulfing a supermassive black hole of mass $\MBH$ in the center of a galaxy. Assuming that all stars are of equal mass $M_\star$, a total of $N_{\star,\rm tot} = \MBH/M_\star$ are enclosed within the sphere of influence, \textcolor{black}{where the SMBH dominates the potential}. The radius of influence is related to the velocity dispersion of the galactic bulge, $\sigma_h$ as
\begin{equation}
    \rh = \frac{G\MBH}{\sigma_h^2} \,,
\end{equation}
and by assuming the M-Sigma relation \citep[e.g.,][]{Tremaine_2002,Marconi_2003,McConnell_2013}, one finds
\begin{equation}
    \rh \approx 1.0 \, M_{\bullet,6}^{0.5} \, \rm pc \,,
\end{equation}
where $M_{\bullet,6} = \MBH/(10^6 \, \rm M_\odot)$.

\subsection{Two-body relaxation}
The stellar cluster settles onto an equilibrium configuration on the two-body relaxation timescale. At every radius $r$
\begin{equation}
    \ttb(r) \approx \frac{1}{\Omega_k (r) N_\star(r)} \pfrac{\MBH}{M_\star}^2 \ln{\Lambda}^{-1} \,,
\end{equation}
where $\Omega_k(r) = \sqrt{G\MBH/r^3}$ is the orbital frequency of a circular orbit of radius $r$, $N(r)$ is the number of stars enclosed within a radius $r$ and $\ln{\Lambda}$ is the Coloumb logarithm, $\ln{\Lambda} \approx \ln{\MBH/M_\star}$. The relaxation time at $\rh$ is approximately
\begin{equation}
    \ttb(r_h) \approx \frac{G\MBH^2}{\sigma_h^3 M_\star} \ln{\Lambda}^{-1} \approx 10^9 M_{\bullet,6}^{1.25} \; \rm yr \,,
\end{equation}
where we used again the M-Sigma relation and $M_\star = \rm M_\odot$ was assumed.

For galaxies with $\MBH \lesssim 10^7 \, \rm M_\odot$, $\ttb(\rh)$ is shorter than the Hubble time, and their nuclei are expected to have settled onto an equilibrium configuration, given by the single-mass, zero particle-flux Bahcall-Wolf (BW) cusp, with $n_\star(r) \propto r^{-7/4}$ \citep{BW_76,BW_77}.

A small fraction of the stars whose semi-major axis is $r$ have a pericenter distance, $r_p$ much smaller than $r$. The orbital angular momentum of such stars is smaller by a factor $\sqrt{2r_p/r}$ relative to a star with a circular orbit of the same semi-major axis. Assuming a BW profile, the number of stars scales as $N_\star(r) \propto r^{5/4}$, increasing sufficiently rapidly with $r$ such that scatterings at apocenter dominate the torque applied on stars with eccentric orbits. The angular momentum relaxation time for highly eccentric orbits with $r_p \ll r$ is therefore given by
\begin{equation}
    \ttb^J(r,r_p) \approx \ttb(r) \pfrac{r_p}{r} \,,
\end{equation}
where the above expression is proportional to $(r_p/r)$ rather than $\sqrt{r_p/r}$ owing to the diffusive nature of the uncorrelated two-body relaxation process. \textcolor{black}{The angular momentum relaxation timescale for eccentric orbits is shorter than the energy relaxation times by a factor $\ttb^J(r,r_p)/\ttb^E(r,r_p)\approx (r/r_p)^{1/4}$, assuming a BW profile \citep[e.g.,][]{Sari_Fragione_2019}}.

\subsection{Gravitational-wave emission}
A star on a circular orbit loses both angular momentum and energy to gravitational waves on a similar timescale, given by
\begin{equation}
    \tgw = \frac{E}{|\dot{E}|} = 2\frac{J}{|\dot{J}|} \approx \frac{R_g}{c} \pfrac{r}{R_g}^4 \pfrac{\MBH}{M\star} \,,
\end{equation}
where $R_g = G\MBH/c^2$ is the gravitational radius of the SMBH.

The orbital dissipation rate of stars on highly eccentric orbits due to gravitational wave emission can be approximated as
\begin{equation}
    \tgw^E(r,r_p) \approx \tgw(r) \pfrac{r_p}{r}^{7/2} \,,
\end{equation}
shorter than the GW angular-momentum dissipation timescale by approximately $\tgw^E/\tgw^J \approx (r_p/r)$ \citep{Peters_1964}.

\subsection{Two-body vs. Gravitational wave regime}
The gravitational wave evolution timescales is comparable to the two-body timescale on orbits that satisfy
\begin{equation}
    \tgw^E(r,r_p) \approx \ttb^J(r,r_p) \,,
\end{equation}
namely,
\begin{equation}
    r_p = R_g \pfrac{N_\star(r) \ln{\Lambda}}{N_\star}^{-2/5} \,,
\end{equation}
and for a BW profile
\begin{equation}
    r_p = R_g \pfrac{r}{\rh}^{-1/2} \,,
\end{equation}
where order unity factors were omitted. Note that for roughly circular orbits, where $r_p \lesssim r$, the two timescales are comparable for $r\approx r_{\rm GW}$, defined as
\begin{equation}
    r_{\rm GW} \equiv (R_g^2 \rh)^{1/3} \,,
\end{equation}
\textcolor{black}{with the subscript indicating the transition to a GW dominated regime}.

Put differently, a star with a semi-major axis $r \gg r_{\rm GW}$, evolves primarily due to two-body scatterings as long as its orbit is not too eccentric, $r_p \gg R_g (r/\rh)^{-1/2}$. If $r_p \ll R_g (r/\rh)^{-1/2}$, the orbit circularizes due to GW emission on a timescale shorter than the angular momentum relaxation time, $
\ttb^J (r,r_p)$.

\section{Stellar-EMRI Production} \label{sec:StellarInspiralProduction}

Stars of radius $R_\star$ and mass $M_\star$ experience tidal forces of a magnitude comparable to their self-gravity as they approach distances of order
\begin{equation} \label{eq:r_tidal}
    \rtidal \approx R_\star \pfrac{\MBH}{M_\star}^{1/3} \,,
\end{equation}
from the SMBH. In this section we review two dynamical channels that place stars on orbits with pericenter distance $r_p \approx \rtidal$, where mass is peeled from the star due to strong tides. \textcolor{black}{Note that the pericenter passage time on such orbits is approximately $\sqrt{\rtidal^3/G\MBH} \approx \sqrt{R_\star^3 / GM_\star}$, i.e., similar to the star's dynamical timescale.}

We emphasize the distinction between tidal disruption events (TDEs) and stellar inspirals (or stellar EMRIs) in this context. TDEs typically involve stars on nearly parabolic orbits, originating at the outskirts of the sphere of influence, which undergo complete tidal disruption as they approach a distance of order $\rtidal$ from the SMBH. ``Stellar EMRIs" occur when stars approach the black hole on a tight inspiraling orbit, such that at the onset of mass transfer, the orbit is roughly circular ($e \lesssim 0.5$), with hours-day periods, where a small fraction of the stellar mass is stripped at each pericenter passage. \textcolor{black}{We note that ``repeating TDEs", producing flares separated by much longer timescales (e.g., \citealt{Payne_2021} with $\sim 110 \, \rm d$, \citealt{Liu_2022} with $\sim 200 \rm d$ and \citealt{Wevers_2022b} with $\sim 10^3 \, \rm d$) may represent an intermediate class of events, that like stellar-EMRIs, are long-lived and are not fully destroyed at the onset of mass transfer, but like TDEs, have very eccentric orbits. We briefly address these systems in the discussion section.}

\subsection{Single-single scattering and GW capture}
\begin{figure}
    \centering
    \includegraphics[width=\columnwidth]{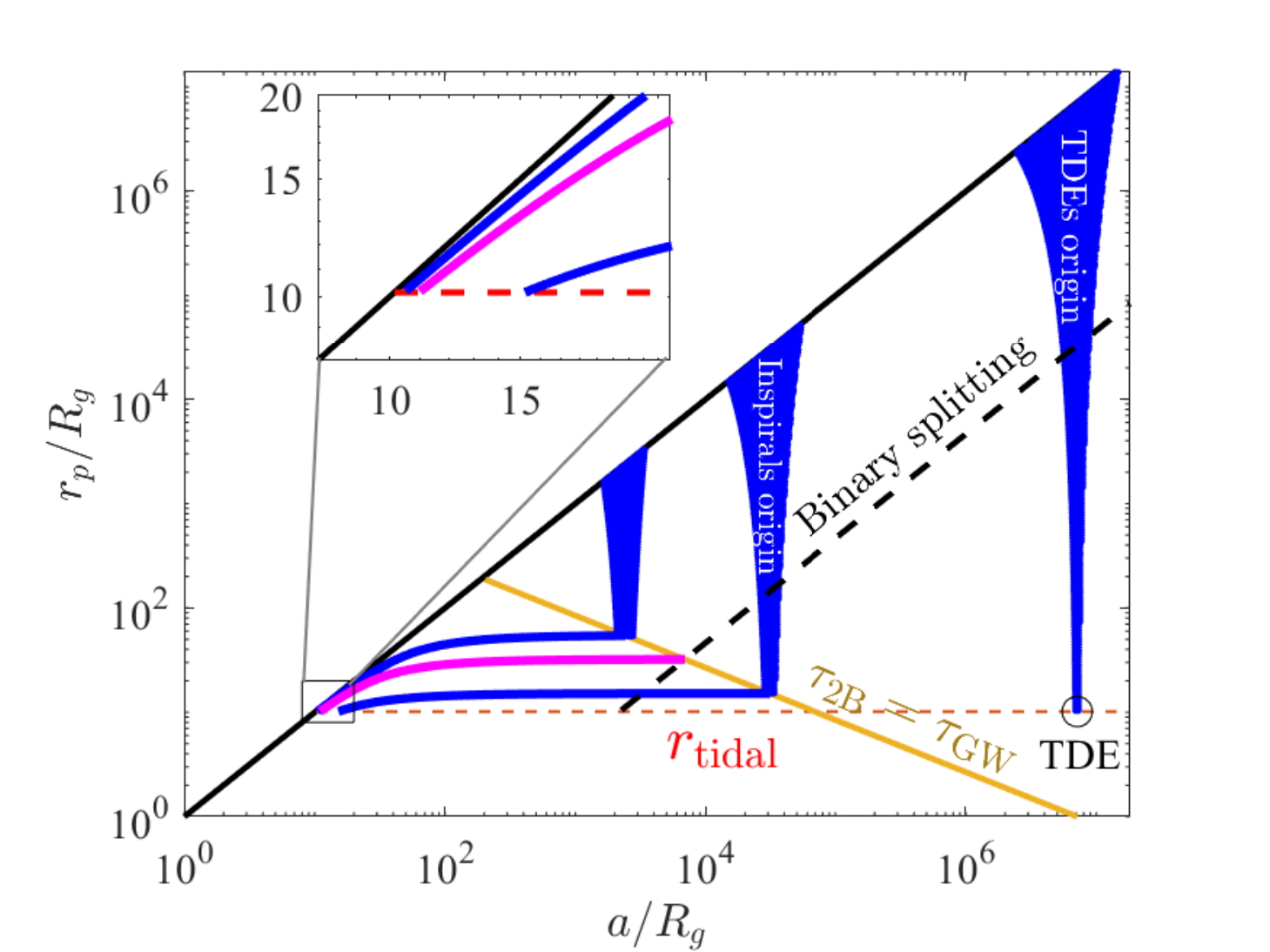}
    \caption{Schematic phase space describing dynamical processes occurring around SMBHs in galactic nuclei - the formation of tidal disruption events (TDEs) and stellar-inspirals. \textcolor{black}{The plot represents the sphere of influence of an SMBH with $\MBH = 10^6 \, \rm M_\odot$}. The vertical axis is the orbital pericenter distance (closely related to the orbital angular momentum), and the horizontal axis is the orbital semi-major axis (related to the orbital energy). The diagonal black solid line corresponds to circular orbits, and the dashed horizontal red line is the tidal radius, where tidal forces exerted by the central supermassive black hole exceed the self gravity of a sun-like star. TDEs are generated as a result of diffusion in angular momentum of stars orbiting the black hole at around the radius of influence, due to two-body encounters (the right blue funnel). Stellar-inspirals are generated primarily by stars on smaller initial radii, with $r_h (R_g/\rtidal)^2$. These stars are stochastically driven to low angular momentum, until circularization through gravitational wave emission becomes dominant (where the blue funnel crosses the yellow line). The star's orbit shrinks until it begins to shed mass onto the black hole, on a tight, mildly eccentric orbit (represented by the stars in the inset). Additionally, Hills' mechanism produces stars on the dashed black line, after which they continue to evolve due to two-body scatterings and GW emission. The magenta track represents the trajectory of the remaining star from a binary split at a radius at which GW emission was marginally more rapid than two-body evolution.}
    \label{fig:PhaseSpaceFigure}
\end{figure}

The combination of two-body scatterings and gravitational wave emission naturally leads to the formation of stellar inspirals, that end up shedding mass to the black hole on mildly eccentric orbits. \textcolor{black}{This process is illustrated in figure \ref{fig:PhaseSpaceFigure}.}

At every radius $r$, stars on semi-circular orbits are scattered to highly eccentric orbits through two-body scatterings at a rate
\begin{multline} \label{eq:BW_2B_rate}
    \mathcal{R}_{\rm 2B}(r) \approx \frac{N_\star(r)}{\ttb(r)} \approx \frac{N_\star^2(r)}{P(r)} \pfrac{M_\star}{\MBH}^2 \approx \\
    \approx \frac{\sigma_h^3}{G\MBH} \pfrac{r}{r_h} \approx 10^{-4} \, \pfrac{r}{r_h} \; M_{\bullet,6}^{-0.25} \; \rm yr^{-1} \,,
\end{multline}
where we assumed a BW profile and the M-Sigma relation. Stars with initial semi-major axis, $r_i$ in the range $r_c < r_i < r_h$, where
\begin{equation}
    r_c \approx \rh (R_g/\rtidal)^2 \,,
\end{equation}
reach $r_p = \rtidal$ through two-body interactions (with $r_i$ essentially unchanged), before GW emission becomes important. These stars therefore comprise the source of TDEs, with their integrated production rate dominated at around $\rh$, namely
\begin{equation}
    \mathcal{R}_{\rm TDE} = \mathcal{R}_{\rm 2B}(\rh) \approx 10^{-4} \; M_{\bullet,6}^{-0.25} \; \rm yr^{-1} \,.
\end{equation}
\textcolor{black}{Formation of TDEs through two-body relaxation is depicted in figure \ref{fig:PhaseSpaceFigure} as the vertical funnel illustrating the phase-space trajectory of such stars, up to the point where $r_p \approx \rtidal$.}

\begin{figure}
    \centering
    \includegraphics[width=\columnwidth]{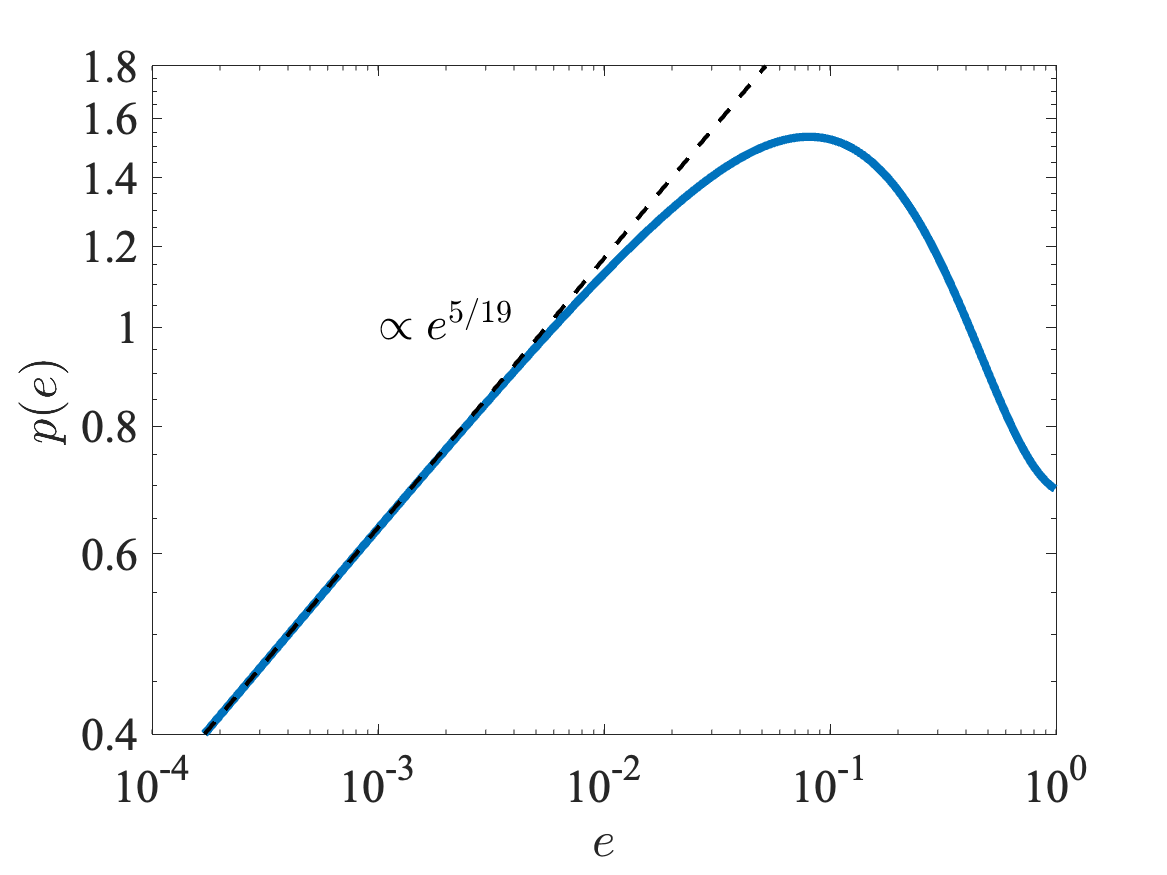}
    \caption{The probability density of the orbital eccentricity at the onset of mass transfer, for systems that circularize due to gravitational wave emission, formed through single-single scatterings. The dashed black line gives the low eccentricity asymptotic, $p(e)\propto e^{5/19}$. The distribution peaks around $e=0.081$.}
    \label{fig:eccentricity_prob_density}
\end{figure}

On the other hand, stars with $r_{\rm GW} < r_i < r_c$ undergo a different evolutionary track. These stars similarly evolve due to two-body angular momentum relaxation, however, before they are disrupted by tides, their orbit begins to circularize due to loss of orbital energy to gravitational wave emission. \textcolor{black}{This process is shown in figure \ref{fig:PhaseSpaceFigure} as the blue tracks that break to the left at the intersection with the yellow separatrix.} The transition to a GW dominated evolution occurs once their pericenter first approaches $r_{p,i}(r_i) = R_g (r_i/\rh)^{-1/2}$ ($\gtrsim \rtidal$). The orbital eccentricity at this phase is very close to 1, with $1-e_i \approx (r_i/r_{\rm GW})^{-3/2} \ll 1$. As the orbit continues to dissipate due to GW emission, the pericenter eventually approaches $\rtidal$, marking the onset of mass transfer. The residual eccentricity $e$ when mass transfer begins satisfies \citep[e.g.,][]{Peters_1964}
\begin{equation}
    g(e) = \frac{\rtidal}{r_{p,i}} = \pfrac{r_i}{r_c}^{1/2} \,,
\end{equation}
where
\begin{equation}
    g(e) = \frac{2e^{12/19}}{1+e}\left[ \frac{304+121e^2}{425} \right]^{870/2299} \,,
\end{equation}
where we approximated $e_i \approx 1$. Since the production rate of stars with initial semi-major axis of order $r_i$ scales as $\mathcal{R}_{\rm 2B}(r_i)\propto r_i$ (equation \ref{eq:BW_2B_rate}), the eccentricity's probability density is given by $p(e) = 2g(e)g'(e)$, plotted in figure \ref{fig:eccentricity_prob_density}. $p(e)$ peaks at around $e_\star = 0.081$ and vanishes as $e^{5/9}$ as $e\to 0$. System with very small eccentricity at the onset of mass transfer, i.e., $e \ll 1$, are unlikely because these would originate from stars with initially small semi-major axis, whose scattering rate is reduced. Very eccentric systems with $1-e \ll 1$ are also rare, because they originate from stars with $r_{p,i}$ very close to $\rtidal$ when gravitational waves take over two-body scatterings, requiring fine tuning. 

The two-body+GW formation channel produces stellar-EMRIs at a rate of
\begin{multline} \label{eq:EMRI_2B_rate}
    \mathcal{R}_{\rm MS-EMRI,single} \approx \mathcal{R}_{\rm 2B}(r_c) \approx \mathcal{R}_{\rm TDE} \pfrac{R_g}{\rtidal}^2 \approx \\
    \approx 10^{-7} \; M_{\bullet,6}^{1.1} \, M_{\star,1}^{-0.9} \; \rm yr^{-1} \,,
\end{multline}
where the M-Sigma and main-sequence relations were assumed.

\subsection{Hills' mechanism} \label{sec:Hills_mechanism}
A fraction $f_b$ of the stars at the sphere of influence are paired in binaries. A binary with semi-major axis $a_b$ is tidally ionized when it approaches a distance
\begin{equation}
    \rtidalb \approx \rtidal \pfrac{a_b}{R_\star} = a_b \pfrac{\MBH}{M_
    \star}^{1/3} \,,
\end{equation}
where we assumed that both binary components are of similar mass $M_\star$. One of the stars escapes as hypervelocity star, whereas the remaining bound star will have a post-disruption semi-major axis of
\begin{equation}
    r_{\rm Hills} \approx \rtidalb \pfrac{\MBH}{M_\star}^{1/3} = a_b \pfrac{\MBH}{M_
    \star}^{2/3} \,,
\end{equation}
independently of the binary's initial semi-major axis around the SMBH (as long as it is much greater than $r_{\rm Hills}$). The remaining star then evolves according to the evolution sketched out in the previous section - it will produce a TDE if $r_{\rm Hills} > r_c$, or as a stellar-EMRI otherwise. \textcolor{black}{This formation channel is demonstrated in figure \ref{fig:PhaseSpaceFigure} as the magenta trajectory, showing the aftermath of a tidally disintegrated binary. The remaining bound star initially has an orbit located on the black dashed line, and then continues to evolve due to GWs/two-body scattering, depending on its position relative to the yellow $\ttb=\tgw$ separatrix. The magenta track corresponds to the case where GWs are (marginally) dominant, and the orbit circularizes and inspirals towards the SMBH, until it intersects the tidal radius (red dashed line) with some small residual eccentricity.}

Note that the Hills' mechanism contributes to the formation of stellar EMRIs, as long as
\begin{equation}
    \rtidal \pfrac{\MBH}{M_\star}^{1/3} < r_c \,,
\end{equation}
or
\begin{equation}
    s_{\rm Hills,GW} \equiv \pfrac{R_\star}{r_{\rm GW}} \pfrac{\MBH}{M_\star}^{4/9} < 1 \,,
\end{equation}
namely, the remaining star following the tidal splitting of a near-contact binary ($a_b \approx R_\star$), has $r_{\rm Hills}$ smaller than the critical semi-major axis for which GWs dominates over two-body relaxation (geometrically, the intersection of the dashed black line and the yellow separatrix lies above $\rtidal$ in figure \ref{fig:PhaseSpaceFigure}). Scaling the left hand-side of the inequality by typical values, we get
\begin{equation}
    s_{\rm Hills,GW} = 0.8 M_{\bullet,6}^{-0.39} M_{\star,1}^{0.35} < 1\,,
\end{equation}
hence, the Hills' mechanism is not expected to contribute to GW driven stellar EMRIs for galaxies with $\MBH \lesssim 10^{5} \rm \; M_\odot$. The lowest possible eccentricity at the onset of mass transfer satisfies $g(e_{\rm Hills, min}) = s_{\rm Hills,GW}$, where the post-binary-disruption eccentricity was approximated to be 1. Since $s_{\rm Hills,GW}$ is never too small ($\ll 1$), and unless finely tuned, never too close to $1$, the minimal residual eccentricity is mild, i.e., $e \sim \mathcal{O}(s_{\rm Hills,GW})$. In what follows, we use $e=0.1$ as our fiducial eccentricity value. We note that our results are quite insensitive to the specific value of $e$ we use, as long as it is not very small or very close to 1. Residual eccentricity of $e = 0.1$ corresponds to stars whose pericenter was $r_{p,i} = \rtidal/g(e) \gtrsim 3 \times \rtidal$ when gravitational waves took over two-body scatterings.

\textcolor{black}{For such systems, tidal heating is not expected to qualitatively affect the evolution we describe (see \citealt{PT_77,Generozov_2018} for related works). We leave a more detailed analysis of the role of the dissipation of tidal energy, during the phase of GW inspiral to a future work}.

Assuming that the distribution of initial binary separations $a_b$ is log-uniform, the Hills mechanism produces stellar-EMRIs at a rate
\begin{multline} \label{eq:EMRI_Hills_rate}
    \mathcal{R}_{\rm MS-EMRI,Hills} \approx f_b\times \mathcal{R}_{\rm TDE} \approx \\
    \approx 10^{-5} \; \pfrac{f_b}{0.1} \MBH^{-0.25} \; \rm yr^{-1} \,,
\end{multline}
up to a small logarithmic correction accounting for the number of decades of $a_b$ that contribute to EMRI production.

Comparing equations \ref{eq:EMRI_2B_rate} and \ref{eq:EMRI_Hills_rate} we see that the Hill's mechanism is typically the dominant source of stellar-EMRI production. This result is due to the fact that at around $r\approx\rh$, which is where most stars are located, single-single scatterings usually result in a TDE before the orbit can circularize through GWs, while sufficiently tight binaries that are scattered at $\rh$ are ionized and leave a star on a tighter orbit that can eventually produce a low-eccentricity stellar-EMRI. 

\textcolor{black}{We discussed the dynamical processes which form both TDEs (with $a \approx \rh$, and $1-e \approx \rtidal/\rh \approx 10^{-5}$) as well as stellar-EMRIs (with $r_p \approx \rtidal \lesssim a$ and $e \approx 0.1$). Stars on orbit that may produce repeating TDEs, with say $P\approx 100 \, \rm d$ and $1-e \approx 10^{-2} $ (reminiscent of \citealt{Payne_2021}) may also form, albeit at lower rates. Referring to figure \ref{fig:eccentricity_prob_density}, the eccentricity distribution of stellar-EMRIs at the onset of mass transfer approaches a constant ($\approx 0.7$) as $e \to 1$, and therefore, systems with $e > 0.99$ comprise roughly $0.7 \%$ of the total stellar-EMRI population formed through single-single scatterings. Unlike the mildly eccentric inspirals previously discussed, highly eccentric systems have $r_p/\rtidal$ that is very close to 1 when GWs become important, possibly causing the star to undergo substantial tidal heating, that may change the subsequent evolution.}

\section{The fate of stellar inspirals} \label{sec:InspiralFate}

As the orbit continues to shrink due to GWs, tides gradually increase relative to the star's self gravity, until mass transfer ensues. Since the remaining eccentricity is non-negligible, the resulting mass transfer is highly variable - mass is easily stripped from the star near pericenter, while near apocenter the star remains mostly intact.

This behavior can also be described by generalizing the common Roche potential formalism, usually defined for circular orbits. In an eccentric orbit, one can consider the `instantaneous' Roche lobe as equivalent to that of an object orbiting the black hole on a circular orbit of the same radius. Since corotation cannot be maintained for an eccentric orbit, the star cannot be stationary in the local corotating frame. Additionally, the time dependent tidal field acting on a (mildly) eccentric orbit excites stellar modes and tidally heats the star, possibly altering the nature of the resulting mass transfer and the stellar structure. Despite the difficulty in incorporating these effects to the standard Roche lobe picture, they likely amount to order-unity corrections to the following basic picture - mass transfer begins when the star first overfills its instantaneous Roche lobe at pericenter, while the star is well contained within its Roche lobe during the rest of the orbit, in between subsequent pericenter passages. The Roche lobe size at pericenter is defined as
\begin{equation} \label{eq:R_RL}
    R_{\rm RL} = r_p \pfrac{M_\star}{\MBH}^{1/3} \,,
\end{equation}
up to some order-unity corrections.

\subsection{Mass transfer stability} \label{sec:MT_stability}
Perhaps the most fundamental aspect determining the star's evolution after the onset of mass transfer is the stability of the resulting mass transfer.

Stability depends on several factors, and in particular on the response of the stellar radius to loss of mass $\varepsilon = d \ln{R_\star}/d\ln{M_\star}$, and on the conservation of orbital angular momentum of mass lost from the star. In the limit $M_\star/\MBH \ll 1$, The stability criterion can be stated as \citep[e.g.,][]{Soberman_1997,Linial_Sari_2017,King_2022}
\begin{equation} \label{eq:MT_stability}
    \frac{5}{6} + \frac{\varepsilon}{2} - \alpha > 0 \,,
\end{equation}
where $\varepsilon = d \ln{R_\star}/d\ln{M_\star}$ and $\alpha$ is the fraction of orbital angular momentum lost from the system due to mass loss, i.e., angular momentum carried by the transferred mass that is ultimately not returned to the orbit. In the above expression, we neglected corrections of order $(M_\star/\MBH)^{1/3} \approx 10^{-2}$.

\textcolor{black}{To evaluate $\varepsilon$, we briefly repeat the arguments and derivations used in section 5.3 of \cite{Linial_Sari_2017}}. The ratio between a star's cooling (Kelvin-Helmholtz) time and the GW evolution timescale at the onset of mass transfer assuming a semi-circular orbits ($r\gtrsim r_p \approx \rtidal$) is given by
\begin{equation}
    \frac{\tgw}{\tkh} \approx \frac{c^5 R_\star^5 L_\star}{G^4 \MBH^{2/3} M_\star^{13/3} } \,,
\end{equation}
where $L\star$ is the star's bolometric luminosity. Assuming power-law main-sequence relations, $R_\star \propto M_\star^{\beta_{\rm R,MS}}$ and $L_\star \propto M_\star^{\beta_{\rm L,MS}}$ with $\beta_R = 0.8$ and $\beta_L = 3.5$ we have
\begin{equation}
    \frac{\tgw}{\tkh} \approx 0.02 \; M_{\bullet,6}^{-2/3} \, \pfrac{M_\star}{M_\odot}^{3.1} \,.
\end{equation}

\textcolor{black}{Since mass loss evolves on a timescale $M_\star/\dot{M}_\star \approx \tgw$ (if mass transfer is stable) or shorter (for unstable mass transfer), we find that for main-sequence stars with $M_\star \lesssim 4 \, M_{\bullet,6}^{0.2}$, such that the condition $\tgw < \tkh$ is satisfied, the star responds \textit{adiabatically} to mass loss. Taking the star's effective adiabatic index to be $\gamma_{\rm ad}$, the corresponding mass radius relation is given by $\varepsilon = (2-\gamma_{\rm ad})/(4-3\gamma_{\rm ad})$ \textcolor{black}{as the star loses mass and evolves away from the main-sequence}. For low mass stars, gas pressure dominates such that $\gamma_{\rm ad} \approx 5/3$ and $\varepsilon = -1/3$, whereas in more massive stars, radiation pressure becomes increasingly important, and $\gamma_{\rm ad}$ approaches $4/3$, such that $\varepsilon$ becomes a large negative number. In summary, for sufficiently low mass stars, we expect $\varepsilon \lesssim -1/3$, corresponding to the adiabatic response of the star obtained for $\tgw<\tkh$.}

The remaining piece of the puzzle is the value of $\alpha$ - the fraction of angular momentum carried by the lost mass that is ultimately transported back to the orbit. Precise determination of $\alpha$ and the stability of the mass transfer may require detailed numerical simulations involving the stellar structure, relativistic orbital motion etc., but within the scope of this paper, we address two important effects which help constrain the value of $\alpha$.

The extreme ratio between $M_\star$ and $\MBH$ implies a small potential difference between the inner and outer Lagrange points and their corresponding equipotential surfaces \citep[e.g.,][]{Linial_Sari_2017}. Since the star is already overfilling its Roche lobe to some extent, stellar material can easily escape through the L2 Lagrange point. The proximity between the L1 and L2 surfaces implies that once the star begins to shed mass towards the black hole, a comparable amount of mass, $f_{L2}\approx 0.5$ leaks outwards from the star, in the opposite direction. Matter lost through L2 follows an orbit which is external to the star's orbit, and the star will likely torque the resulting circumbinary ring to larger radii. This mass loss component certainly carries away orbital angular momentum from the system, contributing roughly $\sim0.5$ to the value of $\alpha$, and possibly even more if the ring is torqued to large radii.

The proximity of the orbit to the black hole's horizon plays an important role in the determination of $\alpha$. The specific orbital angular momentum is given by
\begin{equation}
    J = \sqrt{G\MBH (1+e) \rtidal} \,,
\end{equation}
where $r_p \approx \rtidal$. Taking $e \approx 0.1$, we find
\begin{equation}
    \left. \frac{J_{\rm ISCO}}{J} \right|_{\rm MS} \approx 0.5 \; \MBH^{1/3} \, M_{\star,1}^{-0.4} \,,
\end{equation}
where $J_{\rm ISCO} = 2\sqrt{3} G\MBH/c$ is the specific angular momentum of the innermost stable circular (ISCO) orbit of a Schwartzschild SMBH, and the main-sequeunce relation $\beta_{\rm MS,R}=0.8$ was assumed. Thus, as material is accreted towards the black hole and approaches the ISCO, it still retains some $\sim 50 \%$ of its initial angular momentum, which is added to the SMBH's spin, implying that a significant fraction of the angular momentum is not conserved in the orbit. Combining these results, we approximate $\alpha$ as
\begin{equation}
    \alpha_{\rm L2 + spin-up} \approx f_{\rm L2} + (1-f_{\rm L2}) \frac{J_{\rm ISCO}}{J} \approx 0.75 \,.
\end{equation}

In summary, the combination of the adiabatic radial response of main-sequence stars to mass loss, in combination with the lack of conservation of orbital angular momentum due to black hole spin-up and L2 mass loss imply an unstable mass transfer, i.e., the stability criterion of equation \ref{eq:MT_stability} is violated. 

The above discussion suggests that stability is certainly not robustly achieved in these systems. The rest of the paper deals mostly with implications of unstable mass transfer. For completeness, we refer the reader to \cite{Linial_Sari_2017} for treatment of the evolution of an MS-SMBH system undergoing stable mass transfer on a circular orbit.

\subsection{Mass transfer stability for low-mass white dwarfs}

Mass transfer from white dwarfs on highly eccentric orbits towards SMBHs has been invoked by several authors, who have considered the consequences of stable mass-transfer driven by GW emission \citep{King_2020,Zhao_2021,King_2022}.

We emphasize that the arguments applied to the case of MS-SMBH, become even more severe for WD-SMBH systems. While for main-sequence star $\rtidal/R_g \approx 10-100$, the tidal radius for WDs is just marginally outside the ISCO for the SMBH masses considered in these studies \citep[e.g.,][]{King_2020,Zhao_2021}. This implies that almost all of the material's initial angular momentum is lost to spinning up the black hole, namely - the mass transfer is highly non-conservative. Furthermore, even if these systems are assumed to evolve stably, the implied values of $\dot{M}_{\rm WD}$ are only achieved if the WD overfills its Roche lobe at pericenter to an extent that would lead to substantial loss of mass (and with it, angular momentum) through L2. These two effects imply $\alpha \lesssim 1$, and the stability criterion of equation \ref{eq:MT_stability} cannot be satisfied (here again, $\varepsilon < -1/3$ as for main-sequence stars).

\subsection{Evolution under unstable mass transfer} 
Unstable mass transfer implies that the system enters a runaway evolution, where the amount of mass lost from the star increases at every pericenter passage.

The small amount of mass initially stripped from the star has negligible effect on the stellar structure and on the orbital evolution. The orbit continues to shrink due to GW emission, such that the star overfills its Roche lobe to a greater extent, gradually increasing the orbit-averaged mass loss rate $ \dot{M}_\star $. We define the relative Roche lobe overfilling extent as (see equations \ref{eq:r_tidal} and \ref{eq:R_RL})
\begin{equation} \label{eq:xi_def}
    \xi = \frac{R_\star-R_{\rm RL}}{R_\star} = \frac{\rtidal-r_p}{\rtidal} \,.
\end{equation}
Its time derivative is given by
\begin{multline}
    \dot{\xi} = \left( \frac{\dot{R}_\star}{R_\star} - \frac{\dot{R}_{\rm RL}}{R_{\rm RL}} \right) (1+\xi) = \\
    =\left( (\varepsilon + 5/3 - 2\alpha) \frac{\dot{M}_\star}{M_\star} - \frac{\dot{r}_p}{r_p} \right)(1+\xi) \,.
\end{multline}
In the above expression, $\dot{M}_\star < 0$, $(\varepsilon + 5/3 - 2\alpha) < 0$ (unstable mass transfer), as well as $\dot{r}_p < 0$ (GWs cause the pericenter to shrink), and therefore, $\dot{\xi} >0$. We further note that the (orbit-averaged) mass loss rate is a sensitive function of $\xi$, with
\begin{equation} \label{eq:M_dot_avg}
    \dot{M}_\star \approx - \frac{M_\star}{P} \xi^k \,,
\end{equation}
where $k \approx 3.5-5$ (see derivation in appendix \ref{appendix_B}). This strong dependence is a result of the density stratification near the stellar surface. Using equation \ref{eq:M_dot_avg}, the equation for $\dot{\xi}$ is rewritten as
\begin{equation}
    \dot{\xi} \approx \\
    \left( P^{-1} \xi^k + \tgw^{-1} \right)(1+\xi) \,,
\end{equation}
where we used $\dot{r}_p / r_p \equiv -\tgw^{-1}$, and omitted order unity prefactors.

Thus, when $\xi$ is very small, it increases roughly linearly with time, $\xi \approx t/\tgw$, where $t=0$ marks the onset of mass transfer. However, as it approaches a value of $\xi_{\rm linear/runaway} \approx (P/\tgw)^{1/k}$, changes in the star's radius and the corresponding shrinkage of the Roche lobe due to mass loss begin to dominate over the GW term, causing $\xi$ to evolve in a runaway fashion.

At this phase, the star sheds ever increasing amounts of mass at every pericenter passage, to the extent that GW emission, which was the initial trigger of this process, stops playing a role in the subsequent evolution. The star would continue to evolve in this runaway manner even if the emission of gravitational waves was somehow artificially stopped. The system continues to evolve on ever decreasing timescales, with
\begin{equation}
    \tau_{\rm runaway}(\xi) \approx \frac{\xi}{\dot{\xi}} \approx P \xi^{1-k} \,,
\end{equation}
representing the timescale spent at any value of $\xi$. As $\xi$ approaches 1, the evolution time becomes comparable to the orbital period (and by definition, the star's dynamical time), at which point nearly half the stellar mass is spilled over just a few orbits. The elapsed time from the onset of mass transfer to the final disruption of the star is of order
\begin{equation}
    \tau_{\rm total} \approx \tgw \pfrac{P}{\tgw}^{1/k} \,,
\end{equation}
or roughly
\begin{equation}
    \frac{\tau_{\rm total}}{P} \approx \pfrac{\tgw}{P}^{1-1/k} \approx 10^6 \, M_{\bullet,6}^{-0.5} \, M_\star^{-2.3}\,,
\end{equation}
orbits, where we assumed $k=4$. We note that while the star does inflate and evolve away from its initial main-sequence structure, throughout most of the runaway evolution, when $\xi \ll 1$, the star has only lost a small fraction of its initial mass, $\delta M_\star \approx \xi M_\star$, and therefore its overall structure does not evolve significantly, at least until $\xi \sim \mathcal{O}(1)$.

\subsection{Emission resulting from transferred mass}
The mass transfer from the star to the black hole can fuel bright electromagnetic emission. For a given $\dot{M}_\star$, the system will produce a time-averaged bolometric luminosity
\begin{equation}
    \left< L \right> \approx \left| \dot{M}_\star \right| c^2 \pfrac{R_g}{r_{\rm diss}} \,,
\end{equation}
where $r_{\rm ISCO} \leq r_{\rm diss} \leq \rtidal$ is the radius at which the orbital energy of the infalling material is dissipated and converted to radiation (assuming a sufficiently efficient radiative process). As $\dot{M}_\star$ increases with time, the averaged emission gradually brightens. Normalized by typical values
\begin{equation} \label{eq:L_inspiral}
    \left< L \right>(\xi) \approx 10^{41} \; \pfrac{10 R_g}{r_{\rm diss}} \pfrac{\xi}{10^{-2}}^{4} \, M_{\star,1}^{0.3} \; \rm erg \, s^{-1} \,,
\end{equation}
where the main-sequence mass-radius relation $\beta_{\rm MS,R}=0.8$ has been assumed.

\section{Quasi-Periodic Eruptions from Unstable Mass Transfer from a Main-Sequence Star} \label{sec:QPEs_from_unstable_MT}

Quasi-periodic eruptions (QPEs) are enigmatic high amplitude bursts of X-ray radiation with a recurrence period of a few to several hours, recently discovered near the central supermassive black holes (SMBHs) of a few distant galaxies \citep{Miniutti_2019,Giustini_2019,Arcodia_2021,Chakraborty_2021}. Each QPE flare has a duration that is roughly $5-20 \%$ of the recurrence period, and emits roughly $10^{45}-10^{46} \; \rm ergs$ in soft X-rays. Different models have been proposed in the literature to explain this phenomena. Several authors have invoked stable mass-transfer from a compact star or white dwarf onto a low-mass SMBH \citep[e.g.,][]{King_2020,King_2022,Zhao_2021}. Mass transfer from a main-sequence star has been recently proposed in \cite{Krolik_Linial_2022}, where shocks self-intersecting streams of material have been considered as the source of radiation. Main-sequence stars have also been considered by \cite{Metzger_2021}, who suggested that QPE flares arise from interaction between two counter-rotating main-sequence stars on circular orbits near the tidal radius, shedding an excess of mass at each conjunction of the two stars. \cite{Xian_2021} considered collisions between a pre-existing AGN disk and a star on an inclined orbit to explain the timing of the flares seen in GSN-069 \citep{Miniutti_2019}.

In this section, we will argue that stellar inspirals may naturally explain some of the fundamental observational properties of the known QPE systems.

\subsection{Orbital period and QPE recurrence time}
Stellar EMRIs orbit the SMBH on a mildly eccentric orbits with semi-major axis $\approx \rtidal$, and have an orbital period $P\approx \tdyn$, that depends only on the mean stellar density. Assuming a standard main-sequence stellar structure and $e=0.1$, and by setting $r_p = \rtidal$, the stellar mass and orbital periods are related by $M_\star = 0.9 \, M_\odot \, P_{10}^{1.4}$ where $P_{10} = P/10 \, \rm h$ (see appendix \ref{appendix_A}).

Mass transfer from stellar inspirals varies considerably along an orbit. The resulting emission would be strongly modulated over the orbital period $P$, as long as the dissipation mechanism is capable of rapidly dissipating the orbital energy of the mass transfer stream, and as long as the heated material can readily cool to allow for large variability in the emitted flux \citep[e.g.,][]{Krolik_Linial_2022}. One natural dissipation mechanism capable of producing the high-amplitude, short duty cycle and observed temperature of the QPE flares is due to shocks occurring when streams of mass transfer self-intersect near the ISCO of the SMBH, as discussed in \cite{Krolik_Linial_2022}. On the other hand, the long viscous timescale of a standard accretion disk ($\tau_{\rm visc} \gg P(r_p)$) implies that the high-amplitude, short duty QPE-like flares, cannot originate directly from accretion onto the SMBH, challenging the picture portrayed in \cite{Metzger_2021}. However, in their scenario, mass transfer streams originating from the two counter-rotating stars might mix and have a reduced angular momentum, possibly leading to the formation of an accretion disk with $r < \rtidal$, that may have a sufficiently short viscous time. 
Interpreting the recurrence time of QPEs as the orbital period of stellar inspirals, we find that the range of observed periods $P=2.4-18.5 \rm \, h$, corresponds to stellar masses in the range $0.1-2.2 \, \rm M_\odot$ - which comprises the majority of stellar objects in the galactic center. Our estimates of $M_\star$ for the known QPE systems are summarized in table \ref{table:QPE_Properties}. 

\begin{table*}[t]
\caption{Properties of the known QPE systems. $P$ is the average QPE period, and $\left<L\right>$ is the QPE luminosity averaged over a full period. The Eddington ratio is based on the SMBH mass estimates from \cite{Wevers_2022}. Given these quantities, we estimate $M_\star$ - the mass of the donor main-sequence star, the apparent timescale $\tapp$ and the luminosity evolution time $\tau_{L}$, as well as the Roche lobe overflow extent, $\xi$.}
\begin{tabular}{|c | c c c | c c c c |}
\hline
{\bf Name}	& {$P \; (\rm hr)$} & {$\left< L \right> \; (\rm erg \, s^{-1})$} & {$\frac{\left<L\right>}{L_{\rm Edd}}$} & {$M_\star \; (\rm M_\odot)$} & {$\log_{10} \xi$} & {$\tau_L \; \rm (yr)$} & {$\tau_{\rm app} \; \rm (yr)$} \\ \hline

GSN-069 \citep{Miniutti_2019} & {$9$} & {$5\times 10^{41}$} & {$\sim10^{-2}$} & {$0.8$} & {$-2.4$} & {$14$} & $\sim 8\times 10^3$\\ \hline

RX J1301.9+2747 \citep{Giustini_2019} & {$4.5$} & {$\sim 10^{41}$} & {$10^{-3}-10^{-4}$} & {$0.3$} & {$-2.4$} & {$16$} & $\sim 10^4$\\  \hline

eRO-QPE1 \citep{Arcodia_2021} & {$18.5$} & {$4\times 10^{42}$} &{$10^{-2}-10^{-1}$} & {$2.3$} & {$-1.4$} & {$38$} & $\sim 3\times 10^3$\\ 
\hline

eRO-QPE2 \citep{Arcodia_2021} & {$2.4$} & {$2\times 10^{41}$} & {$10^{-2}-10^{-1}$} & {$0.12$} & {$-2.4$} & {$5$} & {$\sim 3\times 10^3$}\\ 
\hline

2MASXJ0249 \citep{Chakraborty_2021} & {$2.5$} & {$\sim 10^{41}$} & {$10^{-2}$} & {$0.13$} & {$-2.6$} & {$12$} & {$\sim 10^4$}\\ 
\hline

\end{tabular}
\label{table:QPE_Properties}
\end{table*}%

\subsection{QPE energetics}
The observed period-averaged QPE luminosity of roughly $\left< L \right> \approx 10^{41}-10^{42} \, \rm erg \, s^{-1}$ can be explained by an average mass transfer rate $\dot{M}_\star = \left< L \right> (r_{\rm diss}/R_g) / c^2 \approx 10^{-4} \, \rm M_\odot \, yr^{-1}$ onto the SMBH, where $r_{\rm diss}$ was defined as the radius at which the matter's orbital energy is dissipated, and we used a fiducial value of $10 R_g$. Here we assumed that the observed X-ray luminosity is similar to the total bolometric luminosity of the system. This corresponds to an apparent mass-loss timescale, $\tapp = M_\star/\dot{M}_\star \approx 5000 \, \rm yr$ for $M_\star\approx 1 \; \rm M_\odot$ (see table \ref{table:QPE_Properties} for specific values for each system). 
We interpret this implied mass loss rate as originating from a stellar EMRI. From the observerd luminosity, we use equation \ref{eq:L_inspiral} to conclude $\xi = (P/\tapp)^{1/k} \approx 10^{-2}$ (table \ref{table:QPE_Properties}).

We note that the apparent evolution timescale $\tapp$ is several orders of magnitude shorter than the gravitational wave inspiral time of a solar mass main-sequence star orbiting an SMBH at $\rtidal$, \textcolor{black}{$\tgw \approx 10^5 \, M_{\star,1}^{0.9} M_{\bullet,6}^{-2/3} \, \rm yr$} \citep[e.g.,][]{Linial_Sari_2017}. This suggests that the system cannot be undergoing a stable mass transfer driven by GW emission, as that would imply $\tapp \approx \tgw$. We conclude that the system is currently undergoing an \textit{unstable} mass transfer, where the stellar response to mass loss causes the star to expand, therefore $\xi$ to increase, further accelerating the mass loss, in a runaway process. This determination is in agreement with the analysis given in section \ref{sec:MT_stability}, where we have demonstrated that the mass transfer is not conservative enough to be stable.

Our inference of unstable mass transfer has strong consequences on the estimate of the system's remaining lifetime. Significant increase in the accretion rate would occur once the mass of the star changes by only a small fraction $\xi$, rather than by order unity as is in the case of stable mass transfer. Thus, the evolution time is approximately $\tau_L \approx \xi \tapp \approx 10 \, P_{10}^{1.3} \, L_{42}^{-0.7} \, \rm yr$. 
We therefore predict that over the next few years, the known QPEs will continue to brighten, doubling their luminosity within about a decade. These systems will eventually cease to episodically flare and brighten, once they will approach mass transfer rates comparable to the Eddington rate, with $L_{\rm Edd} \approx 10^{44} \, M_{\bullet,6} \; \rm erg \, s^{-1}$, at which point radiation pressure regulates the black hole's accretion rate, and the variable mass transfer will no longer modulate the emitted radiation. \textcolor{black}{The emitted radiation could strongly affect the star itself, potentially enhancing the mass transfer rate, coupling the star's evolution to the radiation that the transfer of mass produces. This effect, discussed in \cite{Krolik_Linial_2022} could significantly affect the long term evolution of the system, possibly regulating the runaway mass-loss. Recently, \cite{Lu_Quataert_2022} also addressed a possible self-regulating effect, where interactions between the star and the accretion flow it produces, ultimately determine the long term steady-sate evolution of the system.}

We emphasize the following hierarchy between the relevant timescales - the star's dynamical time is comparable to the orbital period $P$. The system brightens over a timescale $\tau_L$ corresponding to \textcolor{black}{$\xi^{1-k}\approx 10^6$} orbits. The apparent timescale, $\tapp$ overestimates the actual evolution time by a factor $\xi^{-1}\approx 100$. Finally, the apparent timescale is shorter than $\tgw$ by several orders of magnitude.

\subsection{Rates} \label{sec:rates}
In section \ref{sec:Hills_mechanism} we estimated the stellar-EMRI formation rate through a combination of Hills' mechanism, two-body angular momentum relaxation and GW emission (equation \ref{eq:EMRI_Hills_rate}). The time spent at each logarithmic luminosity range is approximately
\begin{equation}
    \tau_{\rm runaway}(\left< L \right>) \approx P \pfrac{\left< L \right> P}{M_\star (R_g/r_{\rm diss}) c^2}^{1/k - 1} \,,
\end{equation}
where the expression in the parentheses is the per-flare radiated energy relative to the total available energy budget. Normalized by representative values
\begin{equation}
    \tau_{\rm runaway}(\left< L \right>) \approx 10^2 \, P_{10}^{1.3} L_{\rm 41}^{-0.75} \pfrac{r_{\rm diss}}{10 R_g}^{0.75} \; \rm yr \,,
\end{equation}

Combining the formation rate with the expected lifetime implies that the average number of QPE systems per galaxy is of order
\begin{multline}
    N_{\rm QPE} (\left< L \right>) \approx \mathcal{R}_{\rm MS-EMRI,Hills} \times \tau_{\rm runaway} \approx \\
    10^{-4} \; P_{10}^{1.3} L_{\rm 41}^{-0.75} M_{\bullet,6}^{-0.25} \pfrac{r_{\rm diss}}{10 R_g}^{0.75} \pfrac{f_b}{0.1} \,.
\end{multline}

\textcolor{black}{The eROSITA survey detected eRO-QPE1 at $z=0.0505$, covering a volume containing roughly $10^{5}$ galaxies hosting an SMBH with $\MBH \approx 10^6 \, M_\odot$ \citep{Arcodia_2021,Metzger_2021}. The discovery of 2 QPE systems within this volume can therefore be explained by our theoretical abundance estimate.}

Even though the lifetime of brighter systems is shorter, they could be seen to further distances. Hence, in a flux limited survey, bright systems would dominate the number of detections. The exact luminosity upper cutoff is unclear - and is possibly related to the Eddington luminosity of the SMBH. The observed luminosities of the current systems are about 1-3 orders of magnitude smaller than the SMBH Eddington limit \citep{Arcodia_2021,Metzger_2021}, as summarized in table \ref{table:QPE_Properties}. This discrepancy might be explained if either QPEs cannot brighten past their current observed levels and the luminosity stops tracing the mass transfer rate, or alternatively, if mass-loss is self-regulated through the interaction between the star and the accretion flow or the emitted flux \citep[e.g.,][]{Lu_Quataert_2022,Krolik_Linial_2022}.

\section{Summary and discussion} \label{sec:Summary}
We studied systems the formation and evolution of systems containing a main-sequence star orbiting an SMBH on mildly eccentric orbits, episodically shedding mass at every pericenter passage. We highlight the following key properties of our model.

\begin{itemize}
    \item Combination of two-body scatterings, the Hills' mechanism and GW emission produces stars on mildly eccentric orbits around SMBHs, at rates of $\sim 10^{-7}-10^{-5} \; \rm yr^{-1}$.
    \item These stars inspiral towards the SMBH due to GW emission until they begin to shed mass, once $r_p \approx \rtidal$, corresponding to orbital periods of hour-day timescale (essentially dictated by the star's dynamical timescale).
    \item The mass transfer rate is initially governed by the continued GW inspiral and gradually increases with time, but after $\sim 10^6$ orbits, $\sim 10^{-3}$ of the stellar mass has been shed and the stellar response to mass loss begins to dominate the enhancement of mass transfer, over GWs. Mass transfer then increases in a runaway fashion, on ever decreasing timescales.
    \item Mass transfer rates increase from zero to as high as $M_\star/\tdyn \approx 10^3 \, \rm M_\odot \, yr^{-1}$, with all intermediate rates realized along the process. As the system continues to evolve, the time spent around any mass transfer rate becomes shorter, the higher $\dot{M}_\star$ is.
\end{itemize}

We have shown that such systems can naturally produce some of the key features of the known QPE systems, and in particular, their recurrence timescale, emission energy budget, and their rates. We address a few additional aspect and caveats related to the association between stellar-EMRIs and QPEs.

\subsection{Long term evolution of known QPE systems}
\textcolor{black}{A few of the known QPE systems already have years-long observational baselines, which can be compared against the predictions of our model. We emphasize that if mass transfer is self-regulated through interaction between the resulting radiated flux or the accretion flow, the system may not necessarily gradually brighten, and we might underestimate the remaining system lifetime (see end of section \ref{sec:rates}). Nonetheless, we address the properties of the known QPEs in this context.}
The most recently identified QPE candidate, 2MASXJ0249, was discovered in archival XMM-Newton observations, showing two QPE flares in 2006 \citep{Chakraborty_2021}. Based on the 2006 flare data, we estimate the system's evolution timescale as $\tau_L \approx 12 \, \rm yr$, implying that the system is expected to brighten and then stop flaring on this timescale. Indeed, followup observations of the system in 2021, 15 years later, revealed no QPE activity, in agreement with our model's prediction. The period and black hole mass of this system are remarkably similar to those of eRO-QPE2. The current luminosity of eRO-QPE2 is slightly higher than the 2006 luminosity of 2MASXJ0249, implying a shorter evolution time of only 5 years. We predict a similar fate and evolution track for the two systems - we expect eRO-QPE2 to disappear within the next decade, as 2MASXJ0249 did between the 2006 and 2021 observation epochs.

\textcolor{black}{The QPE system with the longest observational baseline is RX J1301.9+2747, which was identified in 2019 using XMM-Newton \citep{Giustini_2019}. Archival observations show that this object was flaring already in the year 2000, with a similar recurrence time of $\sim 4.5 \, \rm hr$. We estimate an evolution time of roughly $\tau_L \approx 16 \, \rm yr$, implying that the system was expected to roughly double in brightness between 2000 and 2019. The observed fluxes are consistent with being constant, given the uncertainties in the spectral fitting performed in \cite{Giustini_2019}. As additional observations are accumulated over the next few years, we anticipate that this system will brighten, until it will eventually approach its Eddington luminosity, at roughly $\sim 10^{44} \, \rm erg \, s^{-1}$, and then disappear.}

\textcolor{black}{The first identified QPE system, GSN-069 was detected by XMM-Newton in 2018. This system has displayed a long term gradual decay in X-ray emission since 2010, possibly indicating a long-lived tidal disruption event \citep{Miniutti_2019}. This gradual long term decay is not explained by our model. The properties of the QPE flares detected in 2018-2019 imply an evolution time of about a decade. However, no QPE activity was detected in a sufficiently long observation performed in 2014. Our model suggests that the QPE was dimmer by a factor of order of unity in 2014, while the gradually evolving X-ray background was brighter by a factor of $\sim 2$. The overall amplitude of the QPE flares with respect to quiescence was therefore a few times lower than in 2018-2019.}

\subsection{Quasi-periodicity.}
QPEs are only quasi-periodic, with a typical scatter in the flare timing of order $\delta t/P \approx 0.1$. One possible origin of this quasi-periodicity may be stellar oscillations. The strong tidal forces exerted on the star can naturally excite different modes of stellar oscillations. Low order radial modes result in periodic changes in the mean stellar density, and as a result - in $\rtidal$. If the stellar oscillations are not in phase with the orbital motion, the peak of mass transfer may occur slightly before or after pericenter. The resulting flares will therefore occur quasi-periodically, with a scatter that depends on the stellar density fluctuations, the orbital eccentricity and on the (average) tidal radius penetration, $\xi$.

\subsection{Formation of narrow-line region.} The accretion of a total mass $\delta M_\star$ onto the black hole is capable of producing a total of $N_{\rm ph} \approx (\delta M_\star/m_e) \alpha^{-2} \approx 10^{63} \, (\delta M_\star/\rm M_\odot)$ hydrogen-ionizing photons, where $m_e$ and $\alpha$ are the electron mass and the fine structure constant. The known QPE systems have shed about $\delta M_\star = \xi M_\star \approx 10^{-2} \, \rm M_\odot$, which could have ionized a total of $\sim 10^5 \, M_{\odot}$ in neutral hydrogen. The elapsed time since the onset of mass transfer is roughly $\Delta t \approx \tgw (P/\tgw)^{0.25} \approx 3500 \, \rm yr$, implying a maximal extent of $\sim 1 \, \rm kpc$ of the ionized region \textcolor{black}{(although most of the radiation is emitted on the much shorter timescale, $\tau_L$, reducing the size of the ionized region)}. Whether or not the QPE emission can sustain a narrow-line region depends on the recombination rate of the ionized gas. If this rate is low enough, our model could potentially explain the recently identified narrow-line regions in the QPE host galaxies \citep{Wevers_2022}, \textcolor{black}{however, more quantitative analysis of this problem is beyond the scope of this paper}.

\subsection{Additional dynamical effects}
In our calculation of the QPE production rate we made several simplifying assumptions regarding galactic nuclei dynamics - we neglected resonant relaxation, stellar collisions and deviations from spherical symmetry. Additionally, we assumed that all stars engulfing the SMBH are of equal mass. More massive objects (i.e., stellar black holes) tend to sink towards the SMBH due to mass segregation \cite[e.g.,][]{AH_09,Linial_Sari_2022}. The concentration of stellar black holes in small radii will tend to scatter low-mass main-sequence stars away from the SMBH, altering the stellar-inspiral rates we report here. 

\begin{acknowledgments}
We would like to thank Nicholas Stone, Riccardo Arcodia, Tsvi Piran, Julian Krolik, Eliot Quataert and Brian Metzger for interesting discussions that contributed to different stages of this work. IL thanks support from the Adams Fellowship, the Rothschild Fellowship and the Gruber Foundation.
\end{acknowledgments}

\appendix

\section{Orbital period of stellar EMRIs} \label{appendix_A}
At the onset of mass transfer, a stellar inspiral has some residual eccentricity $e$ and a pericenter distance
\begin{equation}
    r_p \approx \rtidal = \arl R_\star \left( \MBH/M_\star \right)^{1/3} \,,
\end{equation}
where $\arl \approx 2$ is an order unity prefactor that may weakly depend on $e$. The semi-major axis $a = r_p/(1-e)$, corresponds to an orbital period
\begin{equation} \label{eq:P_e}
    P = 2\pi \sqrt{\frac{a^3}{G \MBH}} = 2\pi \arl^{3/2} (1-e)^{-3/2} \tdyn \,,
\end{equation}
showcasing the fact that at the Roche limit, the pericenter passage time is similar to the star's dynamical timescale, $\tdyn = \sqrt{R_\star^3 / GM_\star}$. Taking the main-sequence star relation $R_\star \propto M_\star^{0.8}$, the stellar mass is related to the period as
\begin{equation} \label{eq:M_from_P}
    M_\star(P) \approx 0.9 \, M_\odot \, \pfrac{P}{10 \, \rm hr}^{1.4} \,,
\end{equation}
where we used the fiducial $e=0.1$.

\section{Mass transfer rate from eccentric inspirals} \label{appendix_B}
Consider a star of mass $M_\star$ and radius $R_\star$ on a marginally mass-shedding orbit $r_p \lesssim \rtidal$. We follow \cite{Linial_Sari_2017} to obtain the mass transfer rate as a function of the Roche-Lobe overfilling extent, $\xi$ (see equation \ref{eq:xi_def}). 

Since the orbit is slightly eccentric, the transfer of mass from the star is varying over the course of an orbital period. The star only overfills its Roche lobe (and transfers mass) during a fraction of the orbit
\begin{equation}
    \frac{\Delta t}{P} = \xi^{1/2} \sqrt{\frac{(1-e)^3}{2 \pi^2 e}} \,,
\end{equation}
valid for $\xi \ll e$. In the opposite regime (not considered here), where $\xi > e$, part of the star is always exceeding the Roche lobe. During this interval, the instantaneous Roche-lobe overfilling varies as
\begin{equation}
    \xi_{\rm i}(t) \approx \xi \left( 1- \pfrac{t}{\Delta t/2}^2 \right) \,,
\end{equation}
where $t=0$ was set to be the pericenter passage time, and the above expression is valid for $-\Delta t/2 < t < \Delta t / 2$.

The instantaneous mass-transfer rate through the L1 nozzle is given by
\begin{equation}
    \dot{M}_{\star,\rm i} (\xi_{\rm i}) \approx \rho v S \,,
\end{equation}
where $\rho$ is the typical flow density, $v$ is the velocity of the flow through the nozzle, and $S$ is the cross section of the flow. Assuming a polytropic stellar structure with index $n$, we estimate
\begin{align}
    \rho \approx K_\rho(n) \frac{M_\star}{R_\star^3} \xi_{\rm i}^n \,, \\
    v \approx c_s = \frac{1}{\sqrt{n}} \sqrt{\frac{GM_\star}{R_\star}} \xi_{\rm i}^{1/2} \,, \\
    S \approx 2\pi R_\star^2 \xi_{\rm i} \,,
\end{align}
where $K_\rho(n)$ is a dimensionless prefactor \citep[see appendix B of][]{Linial_Fuller_Sari_2021} and $c_s$ is the material's initial sound speed. All the quantities were evaluated on the L1 Roche equipotential surface, located beneath the stellar surface at pericenter, at depth $\xi R_\star$. Put together
\begin{equation}
    \dot{M}_{\star,\rm i} (\xi_{\rm i}) \approx K_\rho(n) \frac{2\pi}{\sqrt{n}} \frac{M_\star}{\tdyn} \xi_i^{n+3/2} \,,
\end{equation}
where $\tdyn = \sqrt{R_\star^3 / GM_\star}$.

The \textit{orbit averaged} mass transfer rate is given by
\begin{equation} 
    \dot{M}_\star = \frac{1}{P} \int_{-P/2}^{P/2} \dot{M}_{\star,\rm i}(\xi_{\rm i}(t)) \, dt = \amt \frac{M_\star}{P} e^{-1/2} \xi^{n+2} \,,
\end{equation}
where $\amt$ is a dimensionless prefactor, which we approximate as $\amt \approx 9$ for $n=3/2$ and $\am \approx 2$ for $n=3$. 

\bibliography{main}{}
\bibliographystyle{aasjournal}

\end{document}